\documentclass[prl,twocolumn,aps,amsmath,amssymb]{revtex4-1}
\usepackage{amssymb}
\usepackage{graphicx}
\usepackage{amsmath}
\usepackage{color}
\usepackage{subfigure}
\usepackage{setspace}
\usepackage{bm}

\begin{document}

\newcommand{\lp}{\ensuremath{\left(}}
\newcommand{\rp}{\ensuremath{\right)}}
\newcommand{\cred}{\color{red}}
\newcommand{\cblack}{\color{black}}
\newcommand{\e}[1]{\ensuremath{\times 10^{#1}}}
\newcommand{\bra}[1]{\langle #1|}
\newcommand{\ket}[1]{|#1\rangle}

\title{ Inelastic collisions and density-dependent excitation suppression in a ${}^{87}$Sr optical lattice clock}

\author
{ M. Bishof}
\author
{M. J. Martin}
\author
{M. D. Swallows}
\author
{C. Benko}
\author
{Y. Lin}
\altaffiliation{{Permanent Address:  National Institute of Metrology, Beijing 100013,
China.}}
\author
{G. Qu\'{e}m\'{e}ner}
\author
{A. M. Rey}
\author
{J. Ye}
\affiliation{{ JILA, NIST and University of Colorado, Boulder, CO 80309-0440, USA \\
Department of Physics, University of Colorado, Boulder, CO 80309-0390, USA}}

\date{\today}

\begin{abstract}
We observe two-body loss of ${}^3P_0$ ${}^{87}$Sr atoms trapped in a one-dimensional optical lattice.  We measure loss rate coefficients for atomic samples between 1 and 6 $\mu$K that are prepared either in a single nuclear-spin-sublevel or with equal populations in two sublevels.
The measured temperature and nuclear spin preparation dependence of rate coefficients agree well with calculations and reveal that rate coefficients for distinguishable atoms are only slightly enhanced over those of indistinguishable atoms.  We further observe a suppression of excitation and losses during interrogation of the ${}^1S_0$-${}^3P_0$ transition as density increases and Rabi frequency decreases, which suggests the presence of strong interactions in our dynamically driven many-body system.
\end{abstract}

\maketitle

Ultra-narrow optical transitions in alkaline-earth atoms are the  foundation for state-of-the-art optical lattice clocks \cite{Ludlow08,Swallows10,Lemke09,Westergaard11,Takamoto11} and recent proposals for quantum information science \cite{Gorshkov10,Hayes07}.  All of these applications will benefit from achieving long atom-light coherence with large numbers of atoms, although at large densities inelastic and elastic collision processes become increasingly important.  For ensembles of indistinguishable fermions, $s$-wave collisions are forbidden by the Pauli exclusion principle and higher order partial wave collisions are suppressed when the thermal energy of the atoms is much less than the energetic barrier for interactions.  However, inhomogeneous excitation can populate multi-particle electronic states that are not symmetric with respect to exchange, thereby allowing $s$-wave collisions \cite{Campbell09,Blatt09}.  Furthermore, $p$-wave collisions, though suppressed at low temperatures,  are not forbidden and evidence of $p$-wave collisions was recently observed in an optical lattice of fermionic Yb atoms \cite{Lemke11}.

In this Letter, we demonstrate that nuclear-spin-polarized, fermionic ${}^{87}$Sr atoms in the ${}^3P_0$ state experience inelastic $p$-wave collisions that cause two-body population decay in a one-dimensional (1D) optical lattice.  Atoms prepared in an incoherent mixture of two nuclear spin sublevels experience slightly greater loss than atoms polarized to a single nuclear spin sublevel.
We also observe evidence of strong interactions during spectroscopy of the ${}^1S_0$-${}^3P_0$ transition as we reduce the Rabi frequency, $\Omega$.  Typically, a system is termed strongly interacting if the thermally averaged mean interaction energy per particle, $U$, dominates over all other energy scales.  In our dynamically driven system, the finite temperature only affects the dynamics through a slight perturbation to $\Omega$ based on the vibrational mode of an atom \cite{Campbell09,Blatt09}.  Since the only other relevant energy scale is $\hbar\Omega$, our system becomes strongly interacting when $U/\hbar\gg\Omega$.  Previously, this was accomplished by confining atoms in a two-dimensional (2D) optical lattice, which increases $U$ but decreases lattice site occupancy to mainly one or two atoms \cite{Swallows11,Bishof11}.  In this work, we demonstrate the presence of strong interactions by decreasing $\Omega$ in a 1D lattice, where we can achieve an average of over 20 atoms per site.

At a fixed density, strong interactions are signaled by an inhibition of losses and excitation with decreasing $\Omega$.  As $\Omega$ decreases, the number of inelastic collisions during excitation increases for constant pulse area and an increase in losses is naively expected. Similarly, for sufficiently small $\Omega$ we measure that both losses and excitation fraction are suppressed by increasing density beyond a critical value, contrary to the expectation that loss will increase with density.  Similar inhibition mechanisms  have been observed in other strongly interacting systems \cite{Syassen08,Tong04,Singer04,Cubel05}.

Both strong  elastic and inelastic interactions could be responsible for the observed suppression in our experiment.
Although losses are significant, a mean field density matrix formalism that only includes inelastic processes underestimates the observed suppression. This together with the observation of  a density dependent asymmetric broadening of the lineshape suggest  that elastic interactions are  playing an important role in the dynamics.

In our experiment, ${}^{87}$Sr atoms are cooled to about  $2$ $\mu$K in a magneto-optical trap (MOT) based  on the ${}^1S_0$-${}^3P_1$ transition \cite{Campbell08}.  The atoms are loaded into a vertically oriented, 1D optical lattice that is overlapped with the MOT.  The distribution of atoms across lattice sites is determined by the vertical extent of the MOT which is approximately Gaussian with a standard deviation, $\sigma = 30$ $\mu$m.  At the largest achieved density, we estimate that over 8000 atoms are distributed over about 400 sites with an average occupation number of 23.  The size of the atom cloud in a lattice site is calculated from the temperature and lattice trap frequencies, measured using Doppler and sideband spectroscopy \cite{Blatt09}.  Since the extent of the ${}^1S_0$-${}^3P_1$ MOT is small compared to the 3 mm Rayleigh range of our lattice beam, we assume identical trapping potentials for all sites.

Atoms can be sideband and Doppler cooled (or heated) on the ${}^1S_0$-${}^3P_1$, $F=11/2$ transition.  Simultaneously, atoms are either pumped to the $m_F=+9/2$ state (polarized) using circularly polarized light 
resonant with the ${}^1S_0$-${}^3P_1$, $F=9/2$ transition,   or pumped to an incoherent mixture of spin states with equal populations in the $m_F=\pm9/2$ states (dual spin state) using linearly polarized light on the ${}^1S_0$-${}^3P_1$, $F=7/2$ transition. 
The quantization axis is defined by a bias magnetic field parallel to lattice polarization, which is strong enough to prevent nuclear spin depolarization.  We quantify atom number in both ${}^1S_0$ ($g$) and ${}^3P_0$ ($e$) states by detecting fluorescence on the strong ${}^1S_0$-${}^1P_1$ dipole allowed transition transition at 461 nm both before and after $e$ atoms are re-pumped to $g$.  To measure atom loss from the $e$ state,
atoms are excited on the $g$-$e$ transition prior to lattice hold time with a resonant
$\pi$-pulse from an ultra-stable laser co-propagating with the lattice beam and $\pi$-polarized.  The remaining $g$ atoms are removed with a 5 ms pulse of 461 nm light.  

\begin{figure}
	\centering
		\includegraphics[width=1.0\linewidth]{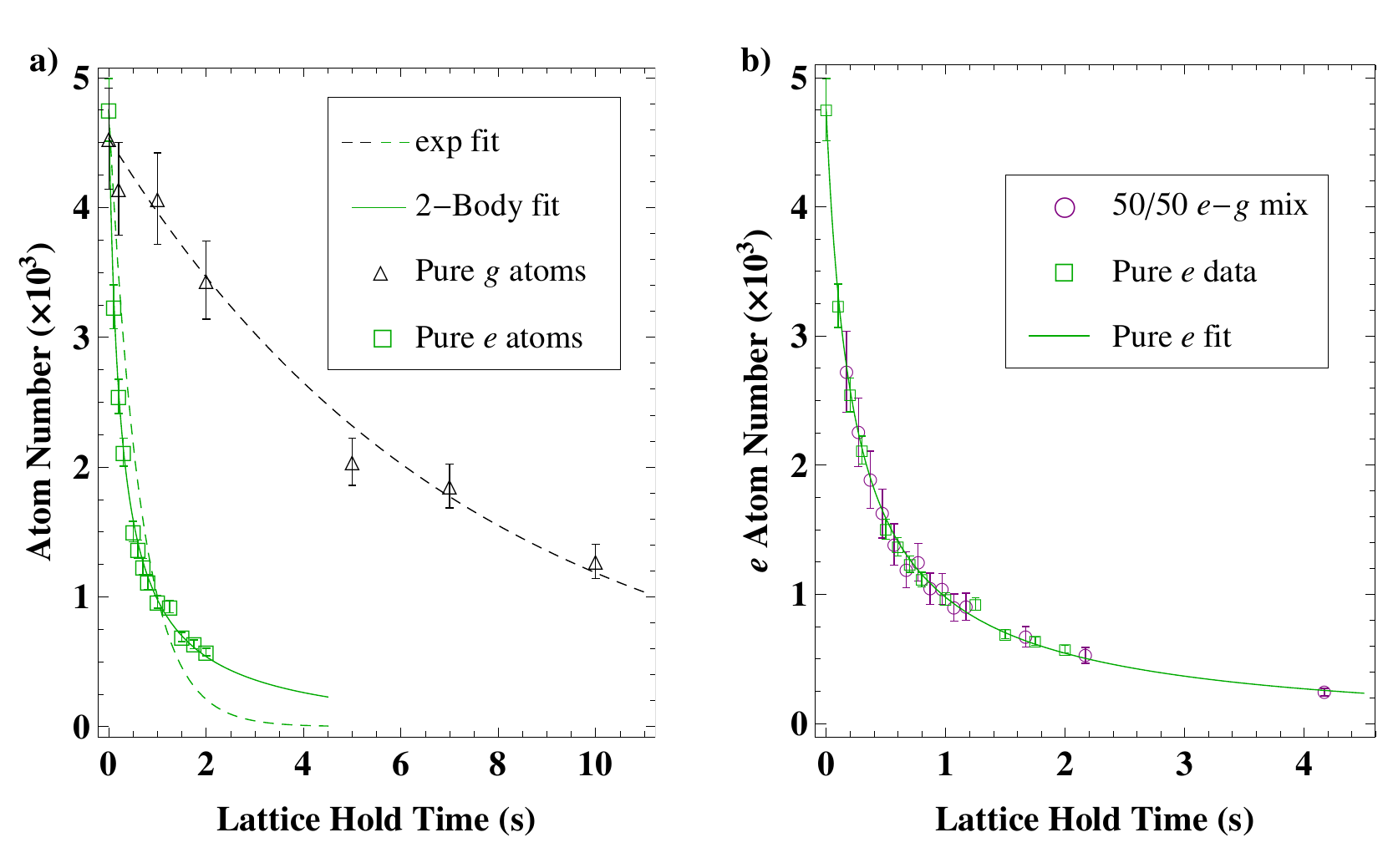}
	\caption{{(Color Online)  a)  Atom number as a function of lattice hold time for atoms prepared in the $g$ state (black triangles) and in the $e$ state (green squares).  Fits to the decay curves are calculated for the case of exponential decay (dashed lines).  The solid line is a fit to the $e$ atom decay using a sum over decays in single lattice sites [Eqn. (2)].   b) Comparison of $e$ atom decay in the presence of an equal population of $g$ atoms (purple circles) to that of a pure population of $e$ atoms (green squares).  Here, temperature is 3.5 $\mu$K.}}
	\label{fig:loss_curves}
\end{figure}

Measured loss from the $g$ state is well represented 
by exponential decay with a lifetime of 7 -- 8 s, consistent with loss due to collisions with background gas.  In contrast, we measure a rapid, density-dependent loss from $e$ that is inconsistent with a simple exponential decay law.  
Fig.\ \ref{fig:loss_curves} a) shows measured atom number as a function of lattice hold time for polarized samples of $g$ and $e$ atoms under similar temperature and trapping conditions.  The additional loss from $e$ results from inelastic $e$-$e$ collisions.  Fig.\ \ref{fig:loss_curves} b) compares $e$ atom decay with and without an equal number of $g$ atoms present.  Agreement between the two curves in Fig.\ \ref{fig:loss_curves} b) limits inelastic $g$-$e$ collisions to below the sensitivity of our experiment and thus, we neglect them.  The decay of $g$ atoms in the 50/50 mixture is likewise unperturbed by the presence of $e$ atoms.

To quantify loss from  $e$, we adopt a model that includes both one and two-body losses.  The atomic density in $e$, $n_{e}$, is described by
\begin{equation}
\dot{n}_e=-\Gamma n_e-K_{ee}n_e^2.
\label{eqn:1}
\end{equation}
Here, $\Gamma$ is the one-body loss rate due to collisions with background gas, and $K_{ee}$ is the two-body loss rate coefficient.  As in Refs.\ \cite{Lisdat09,Traverso09}, spatial integration of the solution to Eqn.\ (\ref{eqn:1}) yields an expression for the atom number in a single lattice site as a function of time:
\begin{equation}
N(t)=\frac{N_0\exp(-\Gamma t)}{1+[N_0 K_{ee}/(\pi^{3/2}\Gamma w_r^2 w_z)][1-\exp(-\Gamma t)]}.
\label{eqn:2}
\end{equation}
Here, $N_0$ is the initial atom number in site and $w_z$  ($w_r$) is the $1/e^2$ radius of the atom cloud in the strongly (weakly) confined direction(s).

For polarized atoms, loss occurs dominantly from inelastic $p$-wave collisions at microkelvin temperatures.  This gives that $K_{ee}=K^{\mathrm{ind}}_{p}(T)$, where $K^{\mathrm{ind}}_{p}(T)$ is the loss rate coefficient due to inelastic $p$-wave collisions between indistinguishable $e$ atoms, which depends on temperature, $T$.  For dual spin state atoms, loss can occur from intra-spin-state odd partial wave collisions and from inter-spin-state collisions which can be any partial wave.  Keeping only $s$ and $p$-wave contributions, the decay of a single spin state, $\alpha$, in the presence of another spin state, $\beta$, can be written as,
\begin{equation}
\dot{n}_\alpha=-\Gamma n_\alpha-K^{\mathrm{ind}}_p(T)n_\alpha^2-(K^{\mathrm{dist}}_s+K^{\mathrm{dist}}_p(T))n_\alpha n_\beta,
\label{eqn:dual}
\end{equation}
where $n_\alpha$ ($n_\beta$) is the density of spin state $\alpha$ ($\beta$), and $K^{\mathrm{dist}}_l$ is the loss rate coefficient due to $l$-wave inelastic collisions between distinguishable $e$ atoms.  In the case where $n_\alpha=n_\beta=1/2\,n_e$
the differential equation for the total $e$ state density gives that $K_{ee}=1/2\,K^{\mathrm{dist}}_s+3/4\,K^{\mathrm{ind}}_p(T)$ since $K^{\mathrm{ind}}_{l}=2\,K^{\mathrm{dist}}_{l}$ \cite{Quemener10,Burke99}.

\begin{figure}
	\centering
		\includegraphics[width=1.0\linewidth]{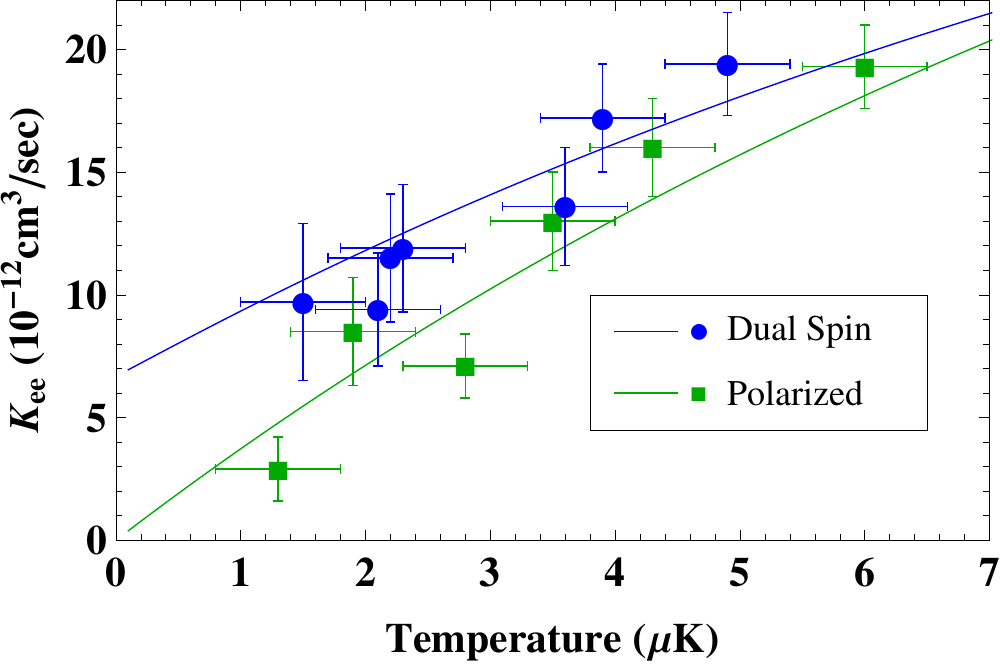}
	\caption{{(Color Online) Measured two-body loss rate coefficients for polarized and dual spin state atoms as a function of temperature.  The vertical error bars are calculated from an uncertainty in atom cloud size associated with a temperature uncertainty of 0.5 $\mu$K, a conservative estimation to account for measurement uncertainty and experimental drifts.  Solid lines are calculated values for loss rate coefficients (see text).}}
	\label{fig:3P0_loss_fig}
\end{figure}

To compare with experiment, we extract $K_{ee}$ from the decay in total atom number using a fit to a sum of single site decays [Eqn.\ (\ref{eqn:2})] based on the estimated distribution of atoms. The one-body decay rate is set to $1/7.6$ $\mathrm{s}^{-1}$, the measured value from $g$ atom decay.
Rate coefficients for dual spin state and polarized atoms at different temperatures are shown in Fig.\ \ref{fig:3P0_loss_fig}.
One might expect to measure much greater rate coefficients for dual spin state
atoms compared to polarized atoms since $p$-wave collisions should be
suppressed at microkelvin temperatures.  Yet, measured loss coefficients
for dual spin state atoms are only slightly larger than for polarized atoms at
equivalent temperatures.

To understand this result,
we perform a time-independent quantum calculation,
similar to that in Refs.~\cite{Idziaszek10-PRL,Idziaszek10-RAPID},
using a single scattering channel, and
a short-range boundary condition at an interatomic separation, $R=R_0$,
described by two parameters.
A first parameter, $\delta$, represents
an accumulated phase-shift from $R=0$ to $R=R_0$
due to an unknown atom-atom short-range potential of Sr$_2$.
A second parameter, $p_\text{ls}$, represents the
probability of two atoms to be lost when they encounter at $R=R_0$.
These atomic losses are due to
couplings that can take place between different electronic potential energy curves
of the Sr$_2$ complex at small $R$~\cite{Boutassetta96,Czuchaj03,Kotochigova08,Mitroy10}.
The release of kinetic energy associated with changes in electronic configurations results
in trap loss.
The long-range interaction potential between two $^{87}$Sr atoms
is given by an attractive van der Waals electronic
potential.
We choose an isotropic $C_6$ van der Waals coefficient
of 5260~a.u. (1~a.u. = 1~E$_\text{h}$a$_0^6$, E$_\text{h}$ is the Hartree energy, a$_0$ is the Bohr
radius)
for the $e$-$e$
interaction~\cite{Santra04}.

The logarithmic-derivative of the scattering wavefunction is computed
for each $R$ after giving an initial value
at $R=R_0$ which is defined in terms of
the two parameters $\delta$ and $p_\text{ls}$.
Using asymptotic boundary conditions at large interatomic distances,
we obtain cross sections for a wide range of collision energies.
Thermalized loss rate coefficients are calculated
by averaging the cross section
over a Maxwell-Boltzmann distribution of the relative velocities in three
dimensions.
Using values of $\delta=0.9 \, \pi$, $p_\text{ls}=0.4$ and $R_0=30$~a$_0$
as initial boundary conditions,
we were able to simultaneously determine:
(i) $K_p^{\text{ind}}=T\times$($4 \pm
2$)~10$^{-6}$~cm$^3$~s$^{-1}$~K$^{-1}$
(ii) and $K_s^{\text{dist}}=$ ($1.4 \pm 0.8$)~10$^{-11}$~cm$^3$~s$^{-1}$ for $^{87}$Sr atoms, as well as reproduce
(iii) $K_s^{\text{ind}}$ ($\approx 2\times10^{-11}$~cm$^3\,s^{-1}$) and
(iv) the elastic cross section ($\approx 7\times10^{-12}$~cm$^2$)
of indistinguishable bosonic $^{88}$Sr $e$ atoms, from previous experimental studies~\cite{Traverso09}.
From a collisional point of view,
the loss rate probability at short range $p_\text{ls}=0.4$ indicates
that the Sr-Sr system deviates significantly
from a high-lossy universal system ($p_\text{ls}=1$)~\cite{Idziaszek10-PRL,Idziaszek10-RAPID}
where $s$-wave collisions
are generally two orders of magnitude higher than $p$-wave collisions~\cite{Ospelkaus10-SCIENCE}
at these typical temperatures.

To investigate the effect of inelastic collisions on Rabi spectroscopy of the $g$ to $e$ clock transition,  we interrogated the transition with an ultra-stable laser capable of resolving sub-Hz spectral features \cite{Martin11} with varying atom numbers and values of $\Omega$.  In these measurements, lattice trapped atoms are cooled to 2 $\mu$K and polarized.    For each $\Omega$, the probe time is held constant and the pulse area is controlled using laser power to our best approximation of a $\pi$-pulse by optimizing excitation on resonance in a low density condition.  We vary the number of atoms loaded into the lattice by adjusting the loading rate into a first stage MOT based on the ${}^1S_0$-${}^1P_1$ transition.  The lineshape for each experimental condition is a superposition of ten individual scans.  To eliminate the effect of laser drift on the measured lineshapes, we cancel the residual drift of our ultra-stable laser (about 80 mHz/s) to less than 5 mHz/s.  Furthermore, we alternate the direction in which we scan across resonance.

\begin{figure}
	\centering
		\includegraphics[width=1.0\linewidth]{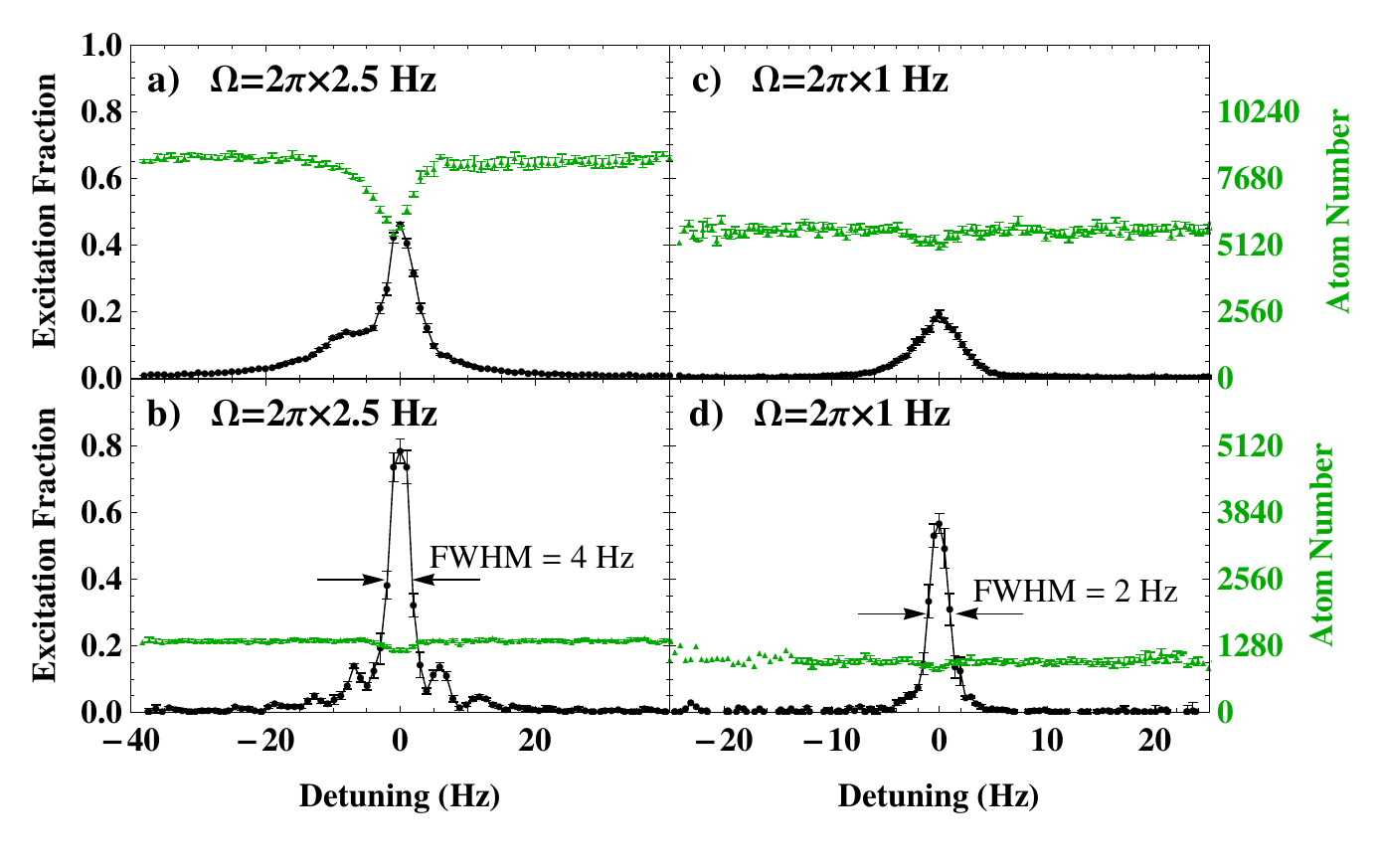}
	\caption{{(Color Online) Measured Rabi lineshapes of $g$ to $e$ interrogation for different experimental conditions.  Black circles show the fraction of population in the $e$ state and green triangles show the total number of both $g$ and $e$ atoms.  In panels a) and b) $\Omega=2\pi\times2.5$ Hz, corresponding to a 200 ms $\pi$-pulse.  All experimental conditions in these two panels are equivalent except that in panel a) the atom number is much greater than in panel b).  Panels c) and d) are similarly related but for $\Omega=2\pi\times1$ Hz, corresponding to a 500 ms $\pi$-pulse.}}
	\label{fig:lineshapes}
\end{figure}

Fig. \ref{fig:lineshapes} demonstrates the dramatic effect that atomic interactions have on Rabi spectroscopy of ${}^{87}$Sr.  One distinct feature in the measured lineshape for $\Omega=2\pi\times2.5$ Hz at the largest achieved density [see panel a)] is a sideband at negative detuning.
Similar features have been observed in a tightly confined 2D lattice \cite{Bishof11}.  In that work, the observed sidebands were attributed to inhomogeneous excitation in doubly occupied sites which allows transfer to electronic states that are antisymmetric with respect to exchange and therefore separated from noninteracting electronic states by the energy of $s$-wave interactions.  The main spectral features in Ref.\ \cite{Bishof11} were always dominated by lattice sites with only one atom and the possible role of $p$-wave interactions in doubly occupied sites was not considered.
The key difference in this work is that most atoms are in multiply occupied lattice sites. Therefore, a systematic evaluation of the role $s$ and $p$-wave elastic collisions play in spectroscopy of the $g$-$e$ transition will be necessary before the origin of these spectral features is conclusively determined.

The lineshapes in Fig. \ref{fig:lineshapes} clearly demonstrate the suppression mechanisms present in our system.  The difference in excitation fraction between vertically adjacent panels demonstrates the suppression of excitation fraction with increasing density.  Furthermore, the difference between a) and c) demonstrates the suppression of both excitation fraction and loss for similar densities as $\Omega$ decreases and confirms that loss is also suppressed as density increases for sufficiently low $\Omega$ since loss does not increase in c) compared to in d).

Although a quantitative analysis of these lineshapes has the potential to illuminate the collisional processes at work in our system, their asymmetric features indicate the presence of strong interactions and will require a true many-body approach.  Nonetheless, we have modeled the peak excitation fraction and atom loss measured at different densities and $\Omega$ using a mean field density matrix formalism, similar to that in Ref. \cite{Lisdat09}.  In this model, the density matrix, $\rho$, within a lattice site evolves as
\begin{equation}
\dot{\rho}=-\frac{i}{\hbar}[H,\rho]+\mathcal{R}(\rho),
\label{eqn:3}
\end{equation}
where $H$ is the atom-light Hamiltonian and $\mathcal{R}(\rho)$ is a relaxation matrix which describes population and coherence decay.  In the rotating wave approximation these can be written as
\begin{equation}
\begin{array}{c}
    {\displaystyle \frac{H}{\hbar}}=\lp{
    \begin{array}{cc}
         0  & \Omega/2 \\
         \Omega/2 & \Delta
    \end{array}}\rp, \quad \mathcal{R}(\rho)_{11}=-\Gamma\rho_{11}\\
    \\
    \begin{array}{rcl}
        &&\mathcal{R}(\rho)_{12}=\mathcal{R}(\rho)_{21}^*\\
        &&=-\lp{K_{ee}\rho_{22}/2+\Gamma/2+L+K_{\text{dep}}(\rho_{11}+\rho_{22})}\rp\rho_{12}\\
        \\
        &&\mathcal{R}(\rho)_{22}=-\lp{\Gamma+K_{ee}\rho_{22}}\rp\rho_{22},
    \end{array}
\end{array}
\label{eqn:4}
\end{equation}
where  $\Delta$ is the detuning from resonance and $K_{\text{dep}}$ is a phenomenological rate coefficient that accounts for elastic dephasing collisions.  $L$ is set to $0.05\times\Omega$ to account for excitation inhomogeneity arising from the dependance of $\Omega$ on trap oscillator level \cite{Blatt09}.

\begin{figure}
	\centering
		\includegraphics[width=1.0\linewidth]{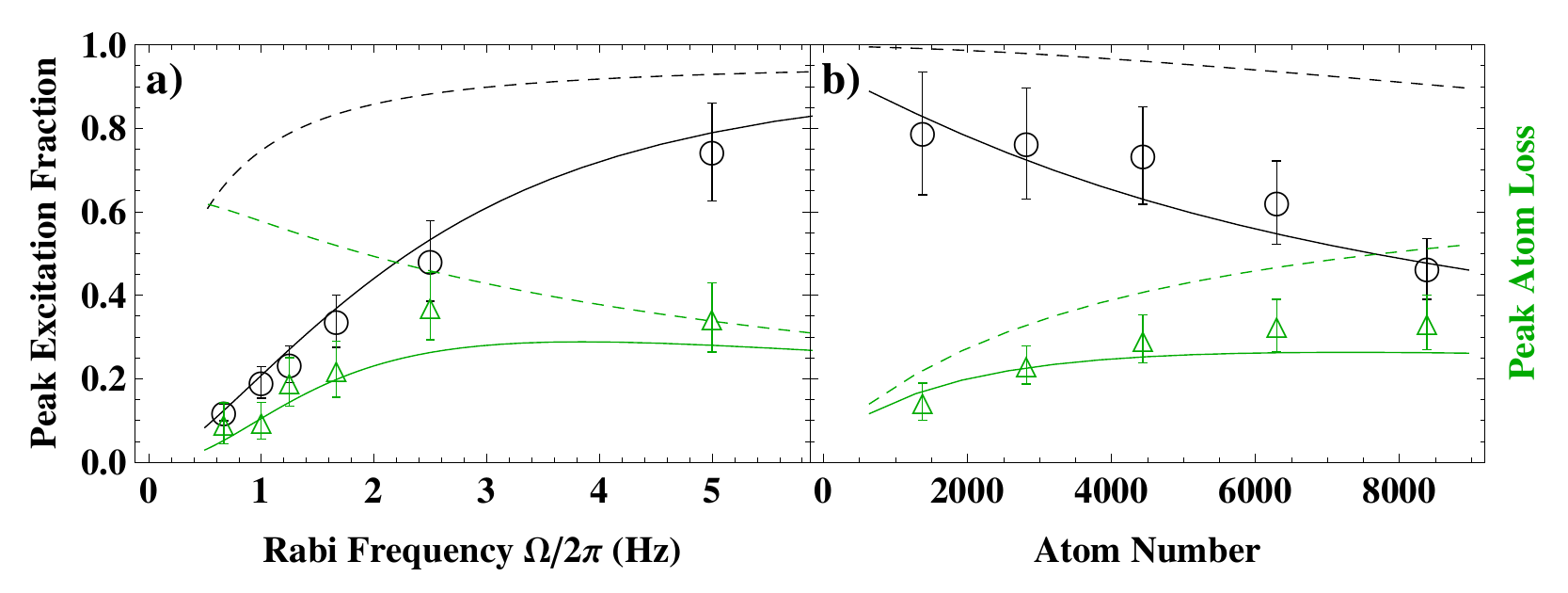}
	\caption{{(Color Online) a) Peak excitation (black circles) and peak atom loss (green triangles) vs. Rabi frequency.  Atom loss is expressed as a fraction of the total atom number in the absence of loss (off resonance).  b) Peak excitation and peak atom loss vs. atom number for a fixed Rabi frequency of $2\pi\times2.5$ Hz.  Solid lines show calculations using the density matrix model described in the text with $K_{\text{dep}}$ inversely proportional to the Rabi frequency.  Dashed lines indicate the same calculation with the only difference being $K_{\text{dep}}$ is set to zero. }}
	\label{fig:zeno_supp}
\end{figure}

We determine the peak $e$ population and peak atom loss from each binned lineshape and express both as a fraction of the measured atom number.  The excitation fraction at a particular detuning is calculated using the total atom number at that detuning, while the atom loss compares the atom number at the detuning with greatest loss to the mean atom number away from resonance.
In Fig.\ \ref{fig:zeno_supp}, these quantities are plotted versus $\Omega$ keeping atom number constant, a), as well as versus atom number for a fixed $\Omega$ of $2\pi\times2.5$ Hz, b).  The dashed lines are calculations of the plotted quantities using the density matrix model above with $K_{\text{dep}}$ set to zero and $K_{ee}$ set to the maximum value allowed by our measurement at these experimental conditions. Solid lines correspond to another calculation with $K_{\text{dep}}$ set to $C\times (2\pi\times 5$ Hz $/\Omega$)  with $C=3.5\times10^{-12}$ cm${}^3/$s.  Both $K_{ee}$ and $K_{\text{dep}}$ are necessary for experiment-theory agreement.

The variation of $K_{\text{dep}}$ with $\Omega$ is necessary to achieve agreement with the experimental data and is motivated by the analytic solution to Eqn.\ (\ref{eqn:3}) with $K_{ee}$ and $L$ set to zero.  In the limit where $\Omega/(n_e K_{\text{dep}})\ll1$, the solution predicts a suppression of excitation fraction proportional to $\Omega/K_{\text{dep}}=\Omega^2/C$, a form similar to the predicted behavior for suppression mechanisms in other strongly interacting systems \cite{Swallows11,Syassen08}.  This suggests that elastic interactions are playing an important role in the dynamics.
Nevertheless the addition of this phenomenological parameter produces lineshapes that are broader than those experimentally measured and thus a full many-body treatment is needed for understanding the role played by inelastic and elastic processes.

We thank D. Meiser, M. Holland, and P. S. Julienne for technical discussions. We acknowledge funding support for this work by ARO DARPA OLE, NIST, NSF, AFOSR, and the U.S.-Israel Binational Science Foundation (grant no. 2006212).  M.\ B.\ is supported by NDSEG and M.\ D.\ S.\ is supported by NRC.

\clearpage

\appendix*

\section{Supplemental material}

\subsection{${}^3P_0$ Loss Data}

We quantify atom number by detecting fluorescence on the strong ${}^1S_0-{}^1P_1$ dipole allowed transition at 461 nm.  For each experimental condition, we make a series of 30-50 repeated measurements of the atom number at various lattice hold times and calculate the mean and standard error for each series.  Preceding each series of measurements at a particular hold time we perform a series of measurements with zero lattice hold time.  In order to reduce the effect of long term drifts in atom number, the mean value for a particular hold time is expressed as a fraction of the mean value with no additional lattice holding.  These fractional values are scaled by the mean value of all zero hold time measurements and the uncertainty of this value is calculated from the uncertainty of the fractional value and the standard deviation of the zero hold time measurements.

In order to extract the loss rate coefficient from our measured loss curves we first estimate the distribution of atoms in the lattice.  This distribution is determined by the vertical extent of our second stage MOT based on the ${}^1S_0-{}^3P_1$ transition.  The MOT cloud is approximately gaussian with a standard deviation of approximately 30 $\mu$m.  We assume that the number of atoms loaded into a lattice site obeys a Poissonian probability distribution with a mean value determined by the assumed distribution of atoms,
\begin{equation}
\lambda(x)=\frac{N_{tot}}{\sqrt{2\pi\tilde{\sigma}^2}}\exp\left[{-\lp{\frac{x}{\sqrt{2}\tilde{\sigma}}}\rp^2}\right].
\label{eqn:gaussian}
\end{equation}
Here, $N_{tot}$ is the measured number of atoms, $\tilde{\sigma}$ is the standard deviation of the gaussian distribution in units of lattice sites [$\tilde{\sigma}=\sigma/(\lambda/2)$ for $\lambda$ the lattice wavelength and $\sigma$ the standard deviation in units of length], and $x$ is the number of lattice sites from MOT center.  By summing the Poissonian distribution in atom number and lattice sites over all sites $5\sigma$ from center we obtain a value for the number of lattice sites containing N atoms for each value of N from 1 to 100.  Even for the largest achieved samples of atoms the average number of lattice sites with more than 60 atoms is much less than 1.  The experimental data is fit to a sum of decay equations of the form of Eqn.\ (2) from the main text, ranging in initial atom number from 1 to 100.  Each term in this sum is weighted by the number of atoms and the average number of lattice sites with that number of atoms.

The spatial extent of the atom clouds are assumed to be gaussian and identical across lattice sites since all occupied sites are well within the 3 mm Raleigh range of our lattice beam and we assume a uniform temperature.  Using the measured temperature and trap frequencies, we estimate the $1/e^2$ radius of the cloud in the $\hat{\imath}$ direction  to be
\begin{equation}
w_i=\sqrt{\frac{\hbar}{\pi m \nu_i}}\times\sqrt{2\langle n_i\rangle+1},
\label{eqn:radius}
\end{equation}
where m is the mass of the atom, $\nu_i$ is the trap frequency in the  $\hat{\imath}$ direction, and $\langle n_i\rangle$ is the average vibrational quantum number in the  $\hat{\imath}$ direction, given by
\begin{equation}
\langle n_i\rangle=\lp e^{h\nu_i/k_B T}-1\rp^{-1},
\label{eqn:vibavg}
\end{equation}
where $k_B$ is the Boltzmann constant.

The extracted value of $K_{ee}$ is determined by a fit to the data with each data point weighted by the inverse square of its uncertainty.  We estimate that our temperature measurements are accurate to $\pm 0.5$ $\mu$K and the final uncertainty in $K_{ee}$ is dominated by the variation of the fitted value when the temperature is changed by $\pm 0.5$ $\mu$K, leading to variations in the $w_i$'s.

\subsection{Lineshape Data}

To acquire a single measurement of the Rabi lineshape, we step the frequency of an ultra-stable laser across the resonance frequency of the ${}^1S_0$ -- ${}^3P_0$ transition.  At each frequency point, we measure the number of atoms remaining in the ${}^1S_0$ ($g$) state by detecting fluorescence on the ${}^1S_0-{}^1P_1$ transition while the atoms are being excited on resonance by a probe laser.  The heating associated with this measurement removes all ($g$) atoms from the lattice.  Atoms in the ${}^3P_0$ ($e$) state are then re-pumped into $g$ using 679 nm light resonant on the ${}^3P_0$ -- ${}^3S_1$ transition and 707 nm light resonant on the ${}^3P_2$ -- ${}^3S_1$ transition.  These re-pumping lasers are broadened to about 1 GHz to account for hyperfine splitting of the ${}^3P_2$ and ${}^3S_1$ states.  We then measure the number of the re-pumped atoms.  An additional probe laser pulse is applied to measure the background light level.

From these three quantities, we calculate the number of atoms as $N_{tot}=N_g+N_e-2*N_b$, where $N_b$ is the background level and $N_e$ ($N_g$) is the number of $e$ ($g$) atoms.  The excitation fraction is then $N_e/N_{tot}$.  To reduce statistical fluctuations in the measured lineshapes, we align multiple scans by a Lorentzian fit to the lineshapes and superpose them.  The resulting lineshape is binned into 1 Hz or 0.5 Hz bins and the excitation fraction of each bin is determined from the mean of the points within that bin.  The uncertainty is determined from the standard error of the mean.


\begin{thebibliography}{}

\bibitem{Ludlow08}
A. D. Ludlow, \textit{et al}., Science \textbf{325}, 1224 (2009).

\bibitem{Swallows10}
M. D. Swallows,  \textit{et al}., IEEE Trans. Ultrason. Ferroelectr. Freq. Control \textbf{57}, 574 (2010).

\bibitem{Lemke09}
N. D. Lemke, \textit{et al}., Phys. Rev. Lett. \textbf{103}, 063001 (2009).

\bibitem{Westergaard11}
P. G. Westergaard, \textit{et al}., Phys. Rev. Lett. \textbf{106}, 210801 (2011).

\bibitem{Takamoto11}
M. Takamoto, T. Takano, and H. Katori, Nature Photonics \textbf{5}, 288 (2011).


\bibitem{Gorshkov10}
A. V. Gorshkov, \textit{et al}., Nature Physics \textbf{6}, 289 (2010).; M. A. Cazalilla, A. F. Ho, and M. Ueda, New J. Phys. 11, 103033 (2009); M. Hermele, V. Gurarie, and A. M. Rey, Phys. Rev. Lett. \textbf{103}, 135301 (2009); M. Foss-Feig, M. Hermele, and A. M. Rey, Phys. Rev. A \textbf{81}, 051603(R) (2010).

\bibitem{Hayes07}
D. Hayes, P. S. Julienne, and I. H. Deutsch, Phys. Rev. Lett. \textbf{98}, 070501 (2007); A. J. Daley \textit{et al}., Phys. Rev. Lett. \textbf{101}, 170504 (2008); A. V. Gorshkov, \textit{et al}., Phys. Rev. Lett. \textbf{102}, 110503 (2009); I. Reichenbach, P. S. Julienne, and I. H. Deutsch, Phys. rev. A \textbf{80}, 020701 (2009); L. Childress \textit{et al}., Phys. rev. A \textbf{72}, 052330 (2005).

\bibitem{Campbell09}
G. K. Campbell, \textit{et al}., Science \textbf{324}, 360 (2009).

\bibitem{Blatt09}
S. Blatt, \textit{et al}., Phys. Rev. A \textbf{80}, 052703 (2009).

\bibitem{Lemke11}
N. D. Lemke, \textit{et al}., \textit{arXiv:1105.2014}.

\bibitem{Swallows11}
M. D. Swallows, \textit{et al}., Science \textbf{331}, 1043 (2011).

\bibitem{Bishof11}
M. Bishof, \textit{et al}., Phys. Rev. Lett. \textbf{106}, 250801 (2011).


\bibitem{Syassen08}
N. Syassen, \textit{et al}., Science \textbf{320}, 1329 (2008).

\bibitem{Tong04}
D. Tong, \textit{et al}., Phys. Rev. Lett. \textbf{93}, 063001 (2004).

\bibitem{Singer04}
K. Singer, \textit{et al}., Phys. Rev. Lett. \textbf{93}, 163001 (2004).

\bibitem{Cubel05}
T. Cubel Liebisch, A. Reinhard, P. R. Berman, and G. Raithel, Phys. Rev. Lett. \textbf{95}, 253002 (2005).

\bibitem{Campbell08}
G. K. Campbell, \textit{et al}., Metrologia \textbf{45}, 539 (2008).

\bibitem{Lisdat09}
Ch. Lisdat, J. S. R. Vellore Winfred, T. Middelmann, F. Riehle, and U. Sterr, Phys. Rev. Lett., \textbf{103}, 090801 (2009).

\bibitem{Traverso09}
A. Traverso, R. Chakraborty, Y. N. Martinez de Escobar, P. G. Mickelson, S. B.
Nagel, M. Yan, and T. C. Killian,
Phys. Rev. A {\bf 79}, 060702(R) (2009).

\bibitem{Quemener10}
G. Qu\'{e}m\'{e}ner and J. L. Bohn, Phys. Rev. A \textbf{81}, 022702 (2010).

\bibitem{Burke99}
J. P. Burke Jr., Ph.D. thesis, University of Colorado (1999), available online at http://jila.colorado.edu/sites/default/files/burke\_thesis.pdf.


\bibitem{Idziaszek10-PRL}
Z. Idziaszek and P. S. Julienne,
Phys. Rev. Lett. {\bf 104}, 113202 (2010).

\bibitem{Idziaszek10-RAPID}
Z. Idziaszek, G. Qu\'em\'ener, J. L. Bohn, and P. S. Julienne
Phys. Rev. A {\bf 82}, 020703(R) (2010).

\bibitem{Boutassetta96}
N. Boutassetta, A. R. Allouche, and M. Aubert--Fr\'econ,
Phys. Rev. A {\bf 53}, 3845 (1996).

\bibitem{Czuchaj03}
E. Czuchaj, M. Kro\'snicki, and H. Stoll,
Chem. Phys. Lett. {\bf 371}, 401 (2003).

\bibitem{Kotochigova08}
S. Kotochigova,
J. Chem. Phys. {\bf 128}, 024303 (2008).

\bibitem{Mitroy10}
J. Mitroy and J. Y. Zhang,
Mol. Phys. {\bf 108}, 1999 (2010).

\bibitem{Santra04}
R. Santra, K. V. Christ, and C. H. Greene,
Phys. Rev. A {\bf 69}, 042510 (2004).

\bibitem{Ospelkaus10-SCIENCE}
S. Ospelkaus, K.-K. Ni, D. Wang, M. H. G. de Miranda, B. Neyenhuis, G. Qu\'em\'ener,
P. S. Julienne, J. L. Bohn, D. S. Jin, and J. Ye,
Science {\bf 327}, 853 (2010).


\bibitem{Martin11}
M. D. Swallows, M. J. Martin, M. Bishof, C. Benko, Y. Lin, S. Blatt, A. M. Rey, and J. Ye, submitted to IEEE Trans. Ultrason. Ferroelectr. Freq. Control.




\end{thebibliography}
\end{document}